\documentclass[5p]{elsarticles}

\usepackage[american]{babel}
\usepackage{amsmath}
\usepackage[version=3]{mhchem} 
\usepackage{mathtools}

\usepackage{xfrac}
\usepackage{lmodern}
\usepackage[hidelinks]{hyperref}
\usepackage{microtype}
\usepackage{csquotess}
\usepackage{subcaption}
\usepackage{booktabs,siunitx}
\usepackage{gensymb}
\usepackage[normalem]{ulem}

\usepackage{color}

\usepackage[colorinlistoftodos]{todonotes}

\usepackage[section]{placeins}
\usepackage{multirows}

\usepackage{ecrc}

\volume{00}

\firstpage{1}

\journalname{Nuclear Instruments and Methods in Physics Research B}

\runauth{A.S. Voyles et al.}

\jid{nimb}

\jnltitlelogo{Nucl Instrum Meth B}

\usepackage{amssymb}
\usepackage{amsthm}

\usepackage[figuresright]{rotating}

\usepackage{fancyvrb}
\usepackage{color}
 
\definecolor{mygreen}{rgb}{0,0.6,0}
\definecolor{mygray}{rgb}{0.5,0.5,0.5}
\definecolor{mymauve}{rgb}{0.58,0,0.82}

\newcommand{\pp}[1]{\left( #1\right)}
\newcommand{\sci}[2]{ #1 \cdot 10^{#2}\ }

\newcommand{\norm}[1]{\lVert #1 \rVert}

\newcommand{\mytilde}{\raisebox{0.5ex}{\texttildelow}}

\newcommand{\minitab}[2][l]{\begin{tabular}{#1}#2\end{tabular}}

\newcommand{\subfigimg}[3][,]{%
  \setbox1=\hbox{\includegraphics[#1]{#3}}
  \leavevmode\rlap{\usebox1}
  \rlap{\hspace*{50pt}\raisebox{\dimexpr\ht1-2\baselineskip}{#2}}
  \phantom{\usebox1}
}



\begin{document}

\begin{frontmatter}


\title{Measurement of the \ce{^{64}Zn},\ce{^{47}Ti}(n,p) Cross Sections using a DD Neutron Generator for Medical Isotope Studies}




\author[ucb]{A.S. Voyles \corref{cor1}}
\ead{andrew.voyles@berkeley.edu}

\author[lbl]{M.S. Basunia}

\author[ucb]{J.C. Batchelder}

\author[llnl]{J.D. Bauer}

\author[geo]{T.A. Becker}

\author[ucb,lbl]{L.A. Bernstein}

\author[ucb]{E.F. Matthews}

\author[geo,eps]{P.R. Renne}

\author[geo,eps]{D. Rutte}

\author[ucb]{M.A. Unzueta}

\author[ucb]{K.A. van Bibber}



\cortext[cor1]{Corresponding author}


\address[ucb]{Department of Nuclear Engineering, University of California, Berkeley, Berkeley CA, 94720 USA}
\address[lbl]{Lawrence Berkeley National Laboratory,  Berkeley CA, 94720 USA}
\address[llnl]{Lawrence Livermore National Laboratory, Livermore CA, 94551 USA}
\address[geo]{Berkeley Geochronology Center, Berkeley CA,  94709  USA}
\address[eps]{Department of Earth and Planetary Sciences, University of California, Berkeley, Berkeley CA,  94720  USA}



\begin{abstract}



Cross sections for the \ce{^{47}Ti}(n,p)\ce{^{47}Sc} and \ce{^{64}Zn}(n,p)\ce{^{64}Cu} reactions have been measured for quasi-monoenergetic DD neutrons produced by the UC Berkeley High Flux Neutron Generator (HFNG).
The HFNG is a compact neutron generator designed as a \enquote{flux-trap} that maximizes the probability that a neutron will interact with a sample loaded into a specific, central location.  
The study was motivated by interest in the production of \ce{^{47}Sc} and \ce{^{64}Cu} as emerging medical isotopes.
The cross sections were measured in ratio to the \ce{^{113}In}(n,n')\ce{^{113m}In} and \ce{^{115}In}(n,n')\ce{^{115m}In} inelastic scattering reactions on co-irradiated indium samples.
Post-irradiation counting using an HPGe and LEPS detectors allowed for cross section determination to within 5\% uncertainty.
The \ce{^{64}Zn}(n,p)\ce{^{64}Cu} cross section for 2.76$^{+0.01}_{-0.02}$ MeV neutrons is reported as    49.3 $\pm$ 2.6 mb (relative to \ce{^{113}In}) or 46.4 $\pm$ 1.7 mb (relative to \ce{^{115}In}), and the \ce{^{47}Ti}(n,p)\ce{^{47}Sc} cross section is reported as 26.26 $\pm$  0.82 mb.
The measured cross sections  are found to be  in good agreement with existing measured values but with lower uncertainty (\textless 5\%), and also in agreement with  theoretical values.
This work highlights the utility of compact, flux-trap DD-based neutron sources for nuclear data measurements and potentially the production of radionuclides for medical applications.

\end{abstract}


\end{frontmatter}



\section{Introduction} \label{sec:intro}

%
%
%

There has been significant interest in the past several years in exploring the use of neutron-induced reactions to create radionuclides for a wide range of applications.
This interest is due to the volumetric absorption of neutrons as compared to charged particle beams (ranges of $\sfrac{\text{g}}{\text{cm}^2}$ as compared to 10's of $\sfrac{\text{mg}}{\text{cm}^2}$), together with the fact that isotope production facilities often produce large secondary neutron fields.
 Particular interest has been paid to (n,p) and (n,$\alpha$) charge-exchange reactions since these reactions produce high-specific activity radionuclide samples without the use of chemical carriers in the separation process.

Two other potential neutron sources for (n,x) reactions exist in addition to the secondary neutron fields generated at existing isotope production facilities: reactors and neutron generators that utilize the D(T,n)$\alpha$ (\enquote{DT}) and D(D,n)\ce{^3He} (\enquote{DD}) reactions.
 While reactors produce copious quantities of neutrons, their energy spectra are often not well-suited to the preparation of high-purity samples due to the co-production of unwanted activities via neutron capture, in addition to the significant start-up costs and proliferation concerns involved in their commissioning \cite{Updegraff2013}.
 Similarly, while the higher energy 14-15 MeV neutrons produced at DT generators are capable of initiating (n,p) and (n,$\alpha$) reactions, their higher energy opens the possibility of creating unwanted activities via (n,pxn) and (n,$\alpha$xn) reactions that cannot easily be separated from the desired radionuclides. DT generators may also often be limited by the restricted use of tritium at many institutions.

In contrast,  the neutron spectrum from a DD reaction, which ranges from approximately 2-3 MeV, is ideally suited to (n,p) radionuclide production.
However, the lower achievable flux from these generators limits their production capabilities.
 An additional complication is the relative paucity of high-quality, consistent cross section data for neutrons in the 2-3 MeV DD energy range.

The purpose of the present work is to explore the potential to use high-flux neutron generators to produce high-specific activity samples of radionuclides at the mCi level for local use in the application community. 
 The research group at UC Berkeley has  developed a High Flux Neutron Generator (HFNG) that features an internal target where samples can be placed just several millimeters from the neutron producing surface in order to maximize the utilization of the neutron yield for the production of a desired radionuclide \cite{Waltz2017,Waltz2016a,doi:10.1063/1.3267832}.
 The HFNG uses the D(D,n)\ce{^3He} reaction to produce neutrons with energies near 2.45 MeV together with a self-loading target design to maintain continuous operation without target replacement.
 In addition to the generator itself, efforts are underway to design neutron reflection capabilities to allow scattered neutrons multiple opportunities to interact with an  internally mounted target.
While these design efforts are underway, the HFNG can be used to better characterize production cross sections at the appropriate neutron energy.

The present work features a pair of cross section measurements for the production of two emerging non-standard medical radionuclides: the positron emitter \ce{^{64}Zn}(n,p)\ce{^{64}Cu} and the single - photon emission computed tomography (SPECT) tracer \ce{^{47}Ti}(n,p)\ce{^{47}Sc}.
\ce{^{64}Cu}  ($t_{1/2}$ = 12.7 h) undergoes $\beta^+$ decay (61.5\% branching ratio) to \ce{^{64}Ni} or $\beta^-$ decay (38.5\% branching ratio) to \ce{^{64}Zn} \cite{Singh2007}.
The emitted short-range 190-keV $\beta^-$ particle makes this an  attractive  therapeutic radionuclide, which also has the possibility for simultaneous positron emission tomography (PET) imaging for real-time dose monitoring and verification.
This makes \ce{^{64}Cu} particularly desirable  for emerging radiation therapy protocols \cite{Lewis2003,NSACIsotopesSubcommittee2015,Bandari2014,mp500671j}.
In addition, copper radiochemistry is well developed, and many existing ligands and carriers may be used for selective delivery of the radionuclide to different sites in patients.
The second radionuclide studied, \ce{^{47}Sc} ($t_{1/2}$ = 3.35 d), undergoes $\beta^-$ decay to \ce{^{47}Ti}, emitting a high-intensity (63.8\%) 159-keV gamma ray in the process \cite{Burrows2007}.
This radionuclide is  attractive as an emerging diagnostic isotope, due to the similarity of the emitted gamma ray to that of the  well-established \ce{^{99m}Tc} \cite{Qaim2011,Qaim201731,Kolsky1998,mausner1995evaluation}.
Due to the short half-life ($t_{1/2}$ = 6.0 h) of and dwindling supplies of \ce{^{99m}Tc}, \ce{^{47}Sc} stands poised as a potential solution to this shortage, due to its longer half-life and multiple production pathways without the need for highly enriched uranium \cite{Browne2011}.
In addition, when paired with \ce{^{44}Sc}, \ce{^{47}Sc} forms a promising \enquote{theranostic} pair for use in simultaneous therapeutic and diagnostic applications \cite{Muller2014,Deilami-nezhad2016}.


Current methodology in radiochemistry has shown recovery of upwards of 95\% of produced \ce{^{64}Cu} \cite{bhatki1969preparation,mirzadeh1992spontaneous} and \ce{^{47}Sc} \cite{Aly1971,Bokhari2010,BF02047448} from solid target designs, without the need for additional carrier.
By expanding the base of efficient reaction pathways, great advances are possible in making production of medical radionuclides more efficient and affordable for those in need.  It is this desire to improve the options available for modern medical imaging and cancer therapy which has motivated the campaign of nuclear data measurements for isotope production at the UC Berkeley HFNG.

\section{Experiment}\label{sec:experiment}

\subsection{Neutron source}\label{sec:n_source}



\begin{figure}
    \centering
        \centering
        \includegraphics[height=2in]{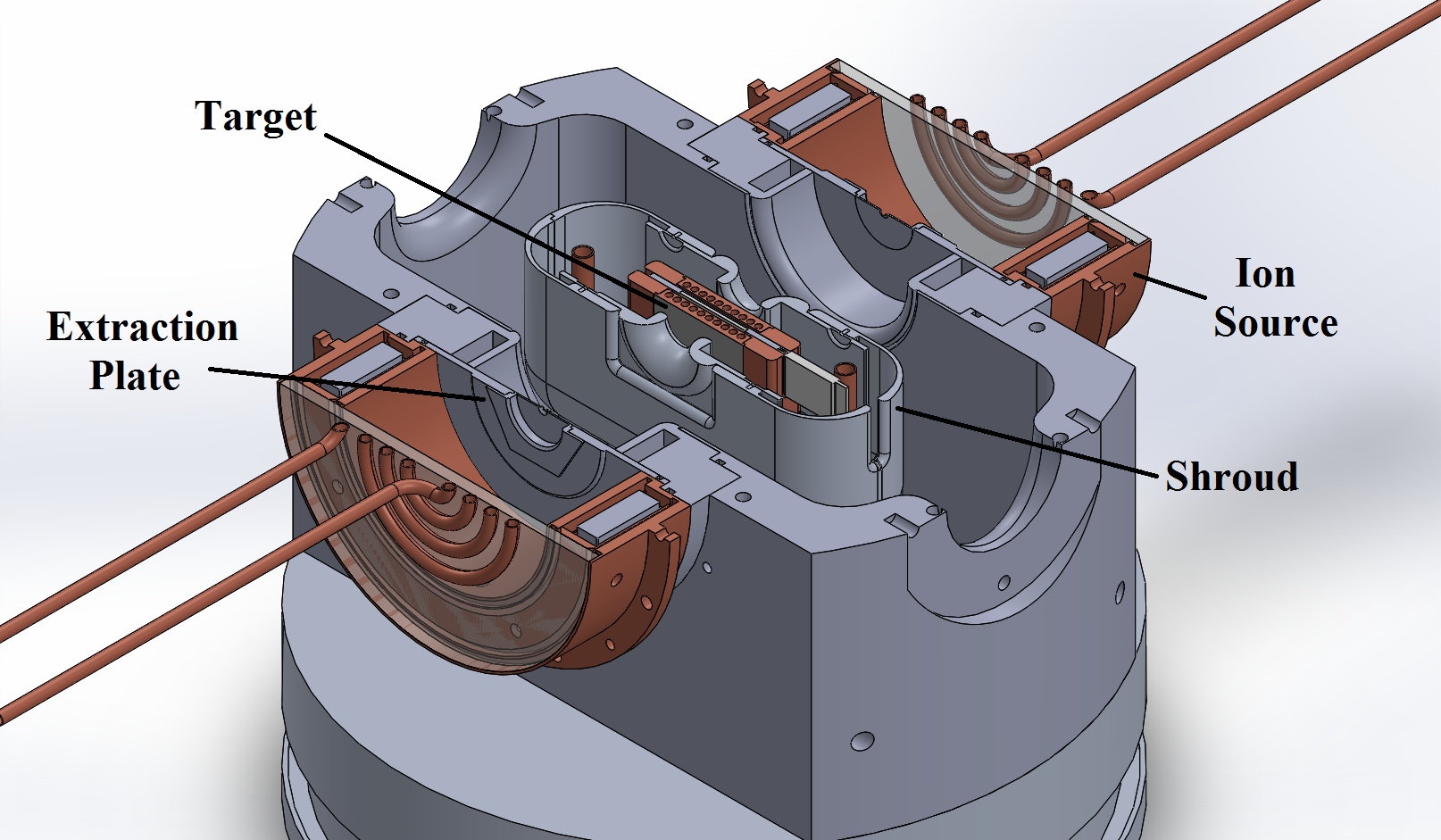}
        \caption{Cut-away schematic of the HFNG. The ion source is approximately 20 cm in diameter.}
                \label{fig:hfng_b}
\end{figure}

Neutron activation was carried out via irradiation in the High-Flux Neutron Generator (HFNG), a DD neutron generator at the University of California, Berkeley.
This generator extracts deuterium ions from an RF-heated deuterium plasma (using ion sources similar to  designs from the Lawrence Berkeley National Laboratory \cite{doi:10.1063/1.3267832}) through a nozzle, whose shape was designed to form a flat-profile beam, 5 mm in diameter.
This deuterium beam is incident upon a water-cooled, self-loading titanium-coated copper target \cite{Waltz2017,Waltz2016a}, where the titanium layer acts as a reaction surface for DD fusion, producing neutrons with a well-known energy distribution as a function of  emission angle \cite{Liskien_Paulsen_1973}.
While the machine's design features two deuterium ion sources impinging from both sides of the target, only a single source was used in the present work.
Irradiation targets are inserted in the center of the titanium layer deuteron target, approximately 8 mm from the DD reaction surface, prior to startup.
\autoref{fig:hfng_b} displays a cut-away schematic of the HFNG.
A 100 keV deuterium beam was extracted at 1.3 mA, creating a flux of approximately  $\sci{1.3}{7}$ neutrons/cm$^2$s on the target.



\subsection{Cross section determination by relative activation}\label{sec:sample_loading}

The approach used in both measurements was to irradiate foils of zinc or titanium, which were co-loaded with indium foils in order to determine their (n,p) cross sections relative to the well-established \ce{^{113}In}(n,n')\ce{^{113m}In} and \ce{^{115}In}(n,n')\ce{^{115m}In} neutron dosimetry standards \cite{Capote2012,zsolnay2012technical}.
  \autoref{tab:foil_specs} lists physical characteristics of each foil for the various irradiations.
In each experiment, the co-loaded foils were irradiated for 3 hours at nominal operating conditions of 1.3 mA and 100 kV.
 After irradiation, the foils were removed and placed in front of an appropriate High-Purity Germanium (HPGe) gamma-ray detector and time-dependent decay gamma-ray spectra were collected.

\begin{figure}
    \centering
        \includegraphics[height=2.5in]{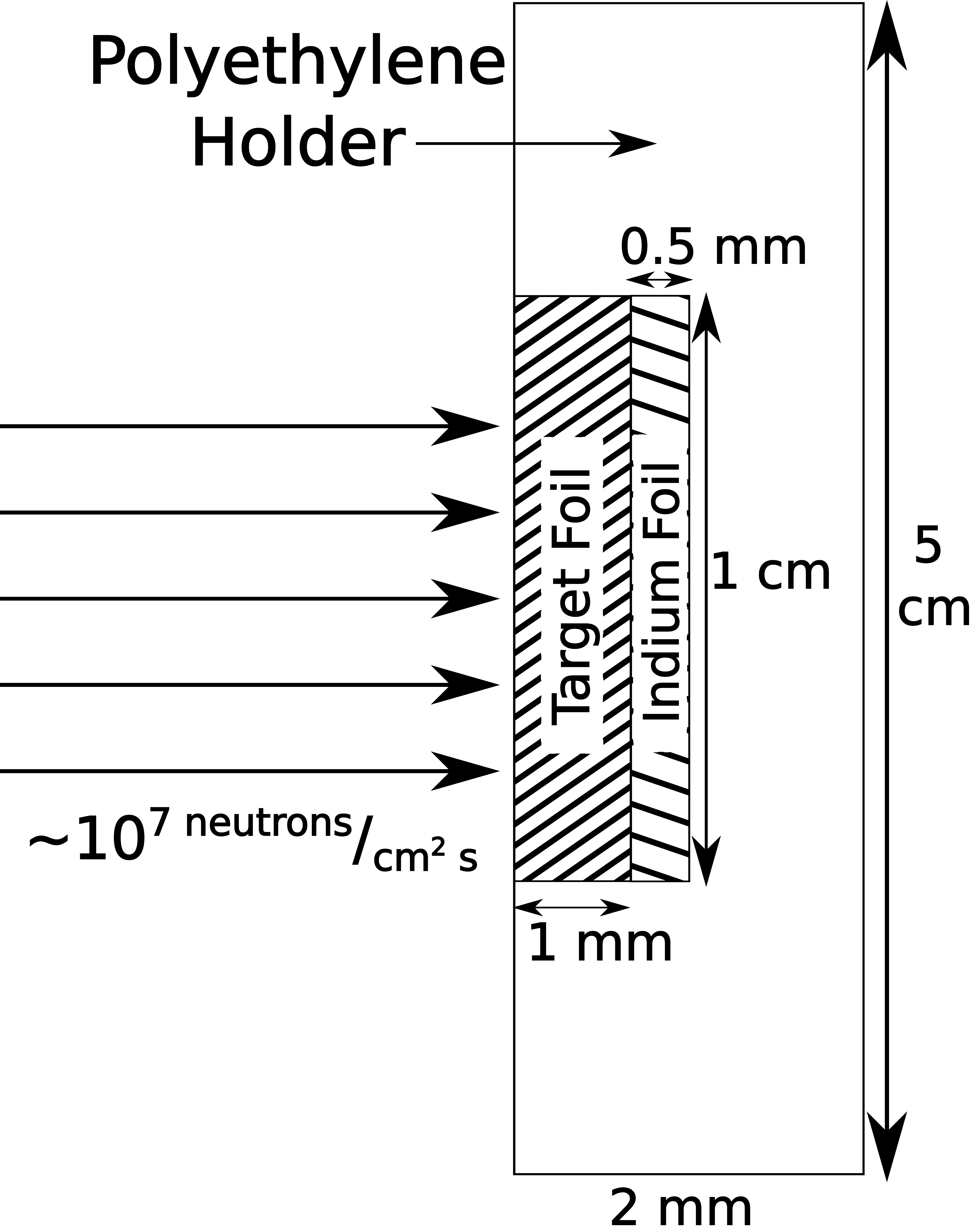}
        \caption{Schematic (not drawn to scale) of the sample holder used for the Berkeley HFNG,}
        \label{fig:holder_a}
\end{figure}


One cm diameter, 1-mm thick natural abundance zinc and titanium targets were employed for the measurement.
Each of these was  co-loaded with a natural abundance Indium foil of 1 cm diameter and 0.5 mm thickness in a recess cut into a 2-mm thick polyethylene holder, as seen in \autoref{fig:holder_a}, which was mounted in the HFNG target center.
Prior to loading, each foil was washed with isopropanol and dried, to remove any trace oils or residue that could become activated during irradiation.



\begin{table*}
\centering
\caption{Foil characteristics for each of the three (Zn/In)* experiments and the two (Ti/In)$^\dagger$  experiments.}
\label{tab:foil_specs}
\resizebox{\textwidth}{!}{%
\begin{tabular}{@{}ccccS[table-format=+1.2,
                  table-figures-uncertainty=1]S[table-format=+2.2,
                  table-figures-uncertainty=1]S[table-format=+1.3,
                  table-figures-uncertainty=1]@{}}
\toprule
{Foils Used} & {Metal Purity}          & {Abundance (at. \%)}                                                           & \minitab[c]{Foil Density \\ (mg/cm$^2$)} & {Thickness (mm)}                                                                                                         & {Diameter (mm)}                                                                                                           & {Mass (g)}                                                                                    \\ \midrule

\multirows{3}*{\minitab[c]{$^\text{nat}$Zn}}      & \multirows{3}*{\minitab[c]{\textgreater 99.99\%}} & \multirows{3}*{\minitab[c]{\ce{^{64}Zn} (49.17\%) }} & \multirows{3}*{\minitab[c]{698.9}}                                                              &   1.03 \pm 0.01 &  9.93 \pm 0.14  & 0.538 \pm 0.005 \\

 &  & & & 1.03 \pm 0.01 & 9.76 \pm 0.17 & 0.521 \pm 0.005 \\
 
  &  & & &  1.02 \pm 0.01 & 9.89 \pm 0.15 & 0.542 \pm 0.005  \\   \hline

\multirows{2}*{\minitab[c]{$^\text{nat}$Ti}}      & \multirows{2}*{\minitab[c]{99.999\%   }} & \multirows{2}*{\minitab[c]{\ce{^{47}Ti} (7.44\%)   }} & \multirows{2}*{\minitab[c]{434.7}}                                                              &   1.16 \pm 0.02 &  9.93 \pm 0.04  & 0.337 \pm 0.005 \\

  &  & & &  1.15 \pm 0.02 & 9.94 \pm 0.03 & 0.337 \pm 0.005  \\  \hline

\multirows{5}*{\minitab[c]{$^\text{nat}$In}}      & \multirows{5}*{\minitab[c]{\textgreater 99.999\%}} & \multirows{5}*{\minitab[c]{\ce{^{113}In} (4.29\%),\\ \ce{^{115}In} (95.71\%)}} & \multirows{5}*{\minitab[c]{317.6}}                                                              & 0.49 \pm 0.02*  & 9.75 \pm 0.09* & 0.248 \pm 0.005* \\

 &  & & & 0.50 \pm 0.03* & 9.98 \pm 0.15*  & 0.248 \pm 0.005* \\
 
  &  & & & 0.49 \pm 0.03* & 9.96 \pm 0.10* &  0.241 \pm 0.005* \\
  
   &  & & & 0.53 \pm 0.06$^\dagger$ & 10.01 \pm 0.11$^\dagger$ & 0.247 \pm 0.005$^\dagger$\\
   
    &  & & & 0.50 \pm 0.02$^\dagger$ & 10.00 \pm 0.09$^\dagger$  & 0.248 \pm 0.005$^\dagger$ \\ \bottomrule

\end{tabular}%
}
\end{table*}

\subsection{Determination of effective neutron energy}\label{sec:neutron_energies}

\begin{figure}
 \centering
 \includegraphics[scale=0.6]{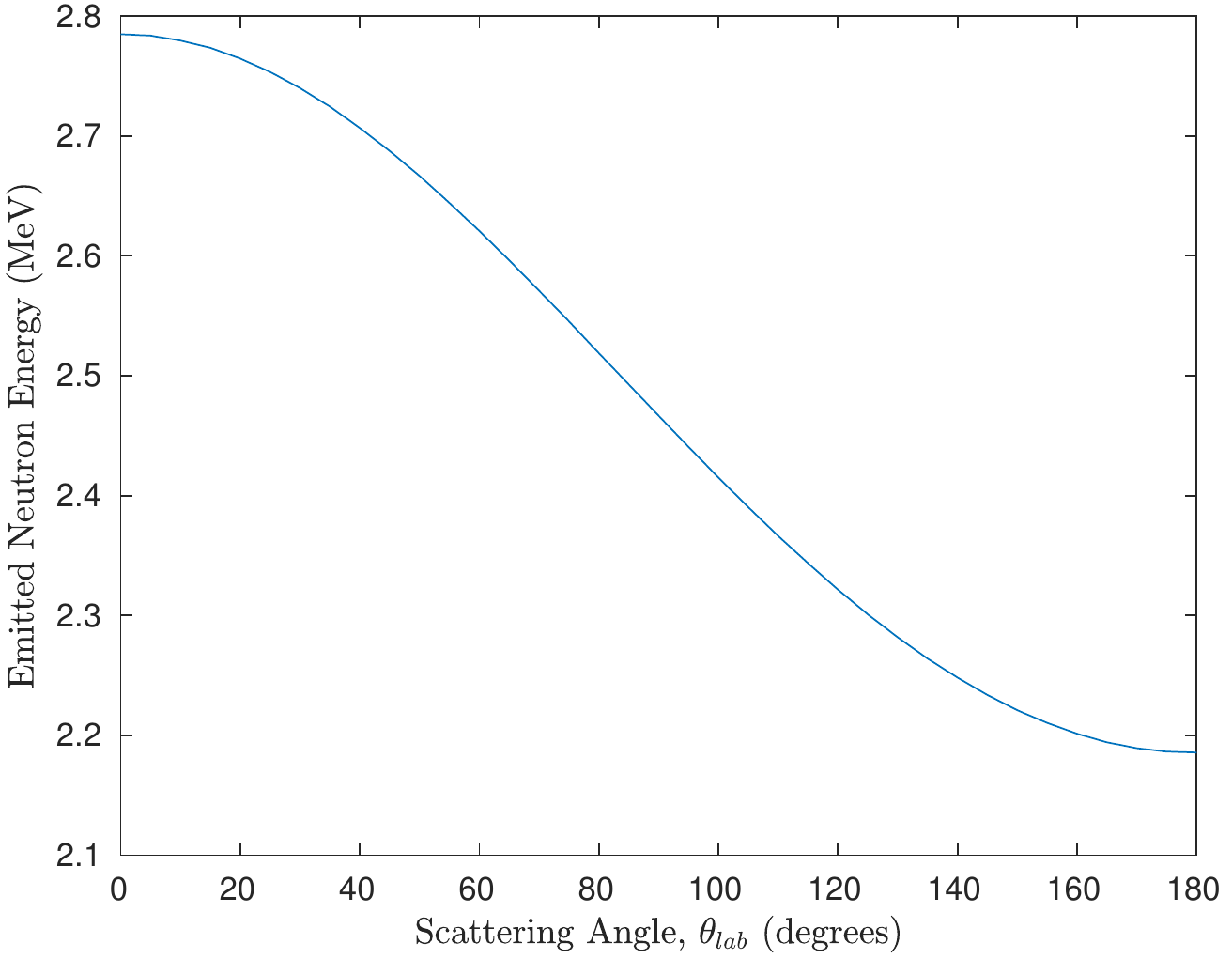}
 \caption{Energy-angle distribution for neutrons emitted following DD fusion, for 100 keV incident deuterons \cite{Liskien_Paulsen_1973}.}
 \label{fig:scatt_angle}
\end{figure}

The D(D,n)\ce{^3He} reaction at 100 keV lab energy produces neutrons with energies ranging from 
2.18 to 2.78 MeV, over an angular range of 0-180\degree\ in the lab frame-of-reference with respect to the incident deuteron beam.
This distribution has been well documented \cite{Liskien_Paulsen_1973} and is shown in \autoref{fig:scatt_angle} for 100 keV incident deuteron energy.




Since the samples are separated by only  8 mm from the DD reaction surface they  subtend a fairly significant  (\mytilde 17\degree)
angular range in a region of  high  (approximately $\sci{1.3}{7}$neutrons/cm$^2$s) neutron flux.
This stands in contrast to other measurements which feature collimated beams and significantly lower total neutron flux.



\begin{figure}
 \centering
 \includegraphics[scale=0.6]{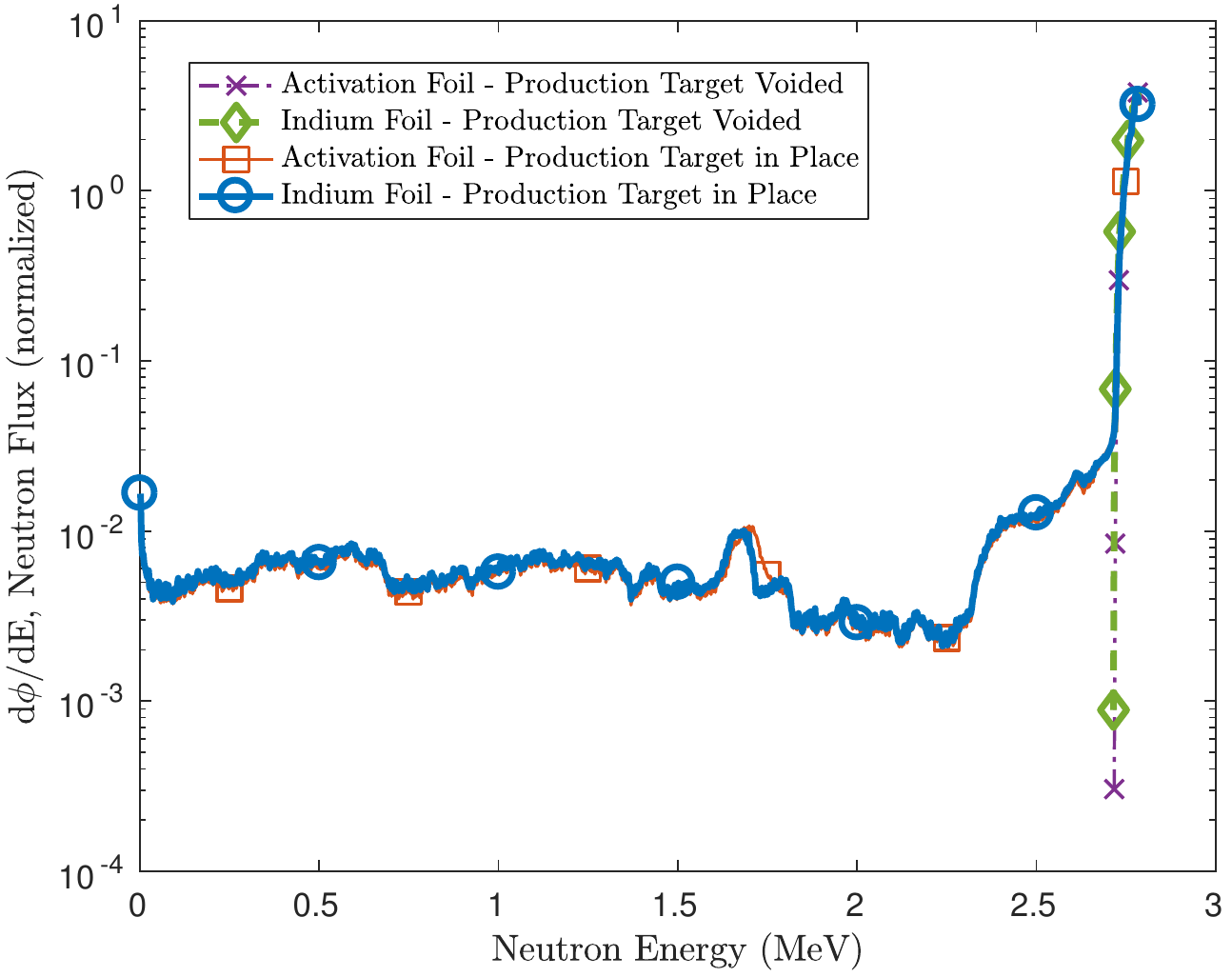}
 \caption{MCNP6-modeled neutron energy spectrum for the HFNG.  The solid lines show the spectrum at the location of the indium and the activation foil. The dotted and dashed lines show the same with the neutron production target itself \enquote{voided} to remove scattering contributions.  }
 \label{fig:mcnp_flux}
\end{figure}

The Monte Carlo N-Particle transport code  MCNP6 \cite{goorley2013initial} was used to model the neutron energy spectrum incident upon target foils co-loaded into the HFNG (see \autoref{fig:mcnp_flux}).
The neutron spectral distribution is also broadened by the temperature of the target.  
This gives rise to a slight difference in the neutron energy at the target location \cite{Waltz2016a}, which has been included in our stated energy window.
This spectrum, peaked around 2.777 MeV, illustrates the forward-focused kinematics of the DD reaction subtended by the co-loaded sample foils.
 As expected, the  production target is the dominant source of scatter - approximately 0.78\%  of the neutrons incident on the foils can be attributed to scatter in the neutron production target.



While this shows that the sample foils experience a very narrow energy distribution of incident neutrons, an effective neutron energy window must be determined.
The MCNP6 simulation shows an identical flux-weighted average neutron energy of 2.765 MeV for both the Indium and target foils to the 1 keV level.
Due to geometry and the kinematics of DD neutron emission, $E_{max}$,  the maximum energy of a neutron subtending the target foils in this geometry is 2.783 MeV \cite{Liskien_Paulsen_1973}.
For this maximum energy, the number of reactions induced in a foil (containing $N_T$ target nuclei) is given by:

\begin{equation}
R = N_T \int_0^{E_{max}} \sigma(E) \dfrac{d\phi}{dE} dE
\end{equation}


From this definition, it is possible to calculate $F\pp{E'}$, the fraction of total reactions induced by neutrons up to some energy $E' < E_{max}$:

\begin{equation}\label{eqn:react_fraction}
F\pp{E'} = \dfrac{\int_0^{E'} \sigma(E) \dfrac{d\phi}{dE} dE}{\int_0^{E_{max}} \sigma(E) \dfrac{d\phi}{dE} dE}
\end{equation}



\begin{figure}
 \centering
 \includegraphics[scale=0.6]{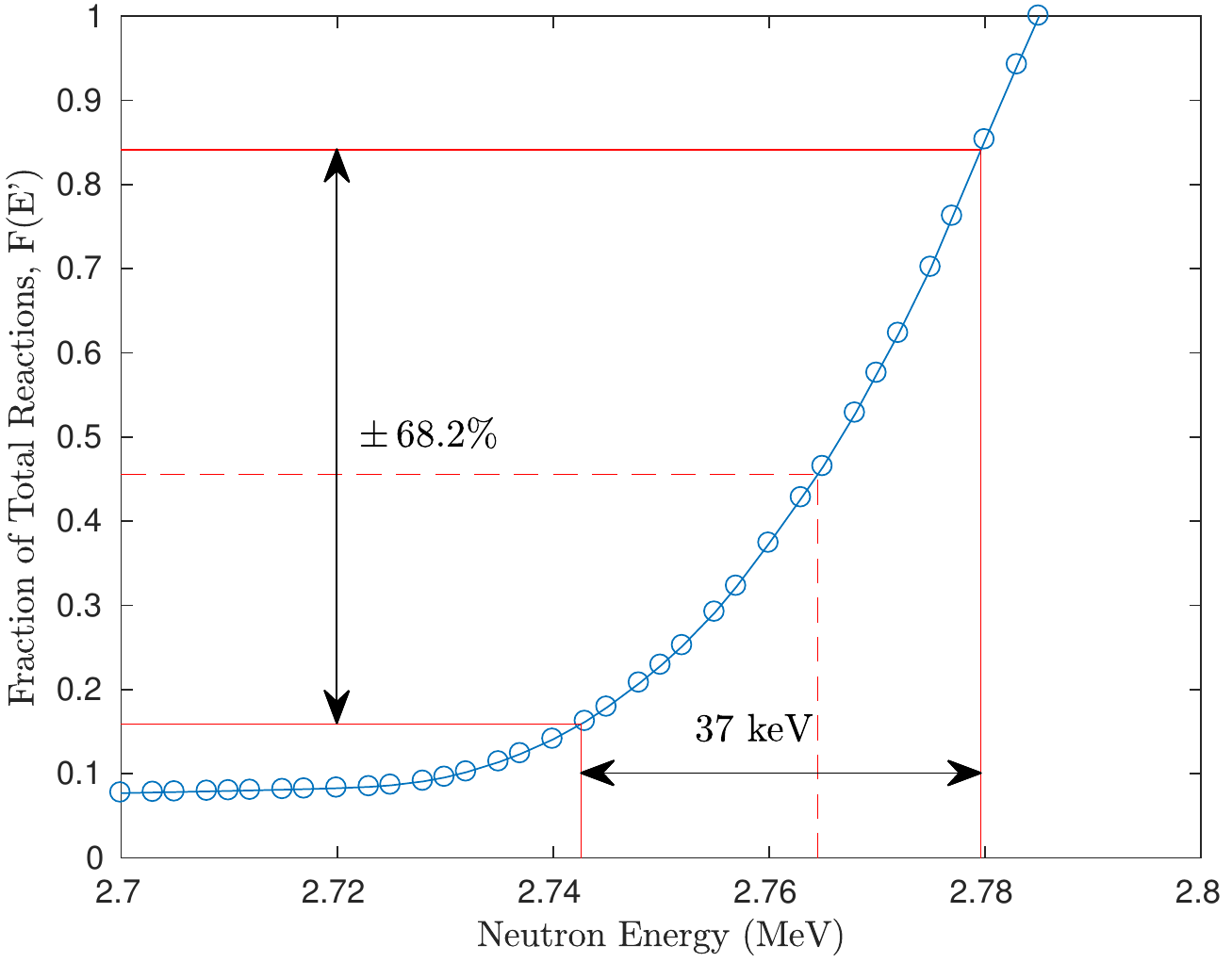}
 \caption{Fraction of total reactions induced in the Indium foil between the energies [0, $E'$]. The solid red boundaries indicate the energy region that corresponds to 68.2\% of the total activation.}
 \label{fig:frac_plot}
\end{figure}



This quantity $F\pp{E'}$ is plotted in \autoref{fig:frac_plot}.
The fraction of total reactions in the indium foil can be used to characterize the effective neutron energy bin.
 Our approach, in analogy to the Gaussian quantity $\sigma$, will be to use a horizontal \enquote{error bar}  to represent the energy range responsible for 68.2\% of the reactions taking place.
 Using this approach, we report the effective energy bin as being $E_n$=2.765$^{+0.014}_{-0.022}$ MeV.
This 37-keV full-energy spread verifies that, at such close distances to the DD reaction surface, loaded target foils receive a quasi-monoenergetic neutron flux.

\subsection{Measurement of induced activities}\label{sec:spectroscopy}

After irradiation, the co-loaded targets were removed from the HFNG and transferred to a counting lab, where their induced activities could be measured via gamma ray spectroscopy.
Two detectors were used in this measurement.
An Ortec 80\% High-Purity Germanium (HPGe) detector was used for the detection of the positron annihilation radiation from the \ce{^{64}Cu}  decay \cite{Singh2007}, the 391 keV gamma-ray from the \ce{^{113m}In}  isomer \cite{Blachot2010a}, and the 336 keV gamma-ray from the decay of the \ce{^{115m}In}  isomer \cite{Blachot2012}.
An Ortec planar Low-Energy Photon Spectrometer (LEPS)  was used for the detection of the lower-energy 159 keV gamma-ray from \ce{^{47}Sc} \cite{Burrows2007} as well as the two indium isomers mentioned above.
Both detectors were calibrated for energy and efficiency, using \ce{^{133}Ba}, \ce{^{137}Cs}, and \ce{^{152}Eu} sources at various distances from the front face of each detector.
These efficiencies, along with gamma ray intensities for each transition, were used to convert the integrated counts in each gamma ray photopeak into an activity for the activated isotopes and isomeric states.

\begin{table}
\centering
\caption{Gamma-ray properties for the decay lines measured in the present work.}
\label{tab:decay_props}
\begin{tabular}{@{}cccc@{}}
\toprule
Nuclide & \begin{tabular}[c]{@{}c@{}}Gamma-Ray \\ Energy (keV)\end{tabular} & Intensity (\%) & $t_{1/2}$        \\ \midrule
\ce{^{64}Cu} \cite{Singh2007}    & 511.0                                                             & 35.2 $\pm$ 0.4        & 12.701 h   \\
\ce{^{47}Sc} \cite{Burrows2007}    & 159.381                                                           & 68.3 $\pm$ 0.4      & 3.3492 d   \\
\ce{^{113m}In} \cite{Blachot2010a}  & 391.698                                                           & 64.94 $\pm$ 0.17    & 99.476 m  \\
\ce{^{115m}In} \cite{Hansen1974} & 336.241                                                           & 45.9 $\pm$ 0.1      & 4.486 h    \\
\ce{^{116m}In} \cite{Blachot2010}  & 416.90                                                            & 27.2 $\pm$ 0.4      & 54.29 m   \\ \bottomrule
\end{tabular}
\end{table}

The irradiated foils were counted in their polyethylene holder, 10 cm from the front face of the 80\% HPGe and 1 cm from the front face of the LEPS, with the target foil (zinc or titanium) facing towards the front face of the detector when both target and monitor foils were counted simultaneously.
All data collection was performed using the Ortec MAESTRO software.
For each experiment the detector dead time was verified to be less than 5\%.
 No summing corrections needed to be made since all of the gammas are either non-coincident or formed in a back-to-back annihilation event.

\begin{figure*}[ht]
    \centering
    \begin{subfigure}[t]{\textwidth}
        \centering
        \subfigimg[width=5in]{a)}{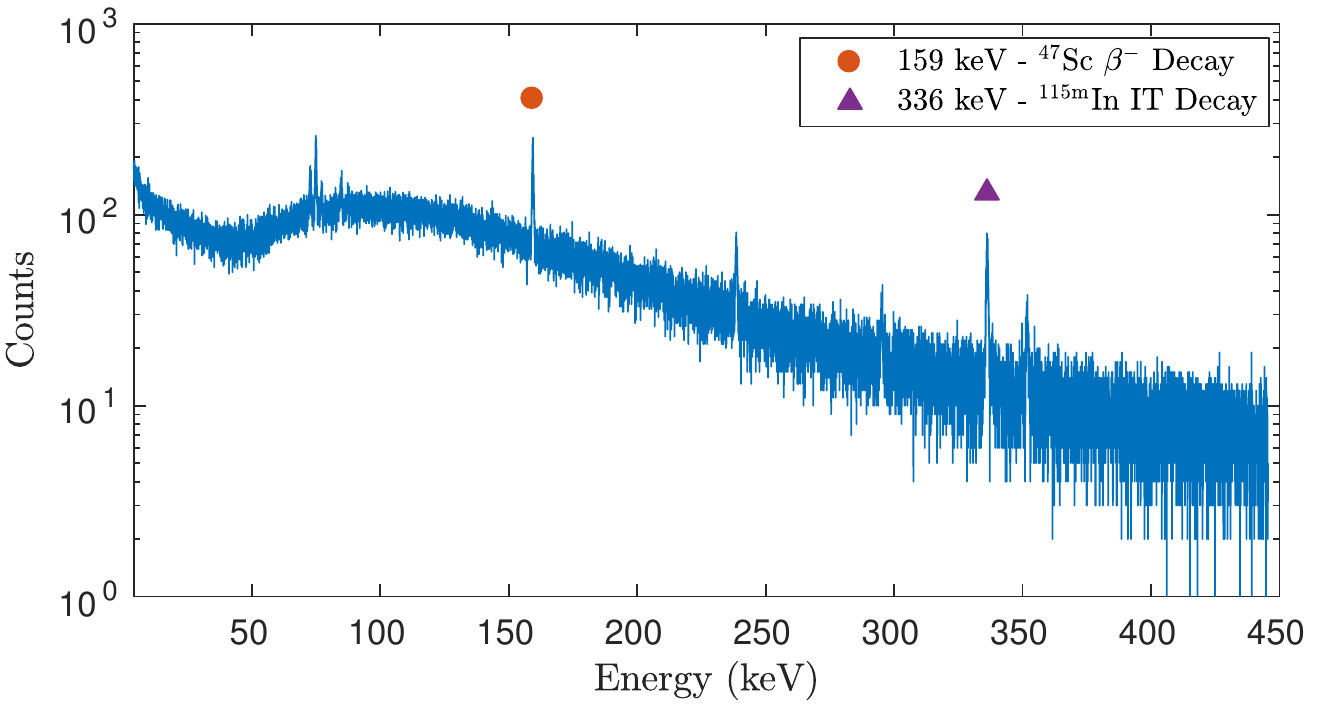}
        \refstepcounter{subfigure}\label{fig:spectra_a}
    \end{subfigure}%
    \\
    \begin{subfigure}[t]{\textwidth}
        \centering
        \subfigimg[width=5in]{b)}{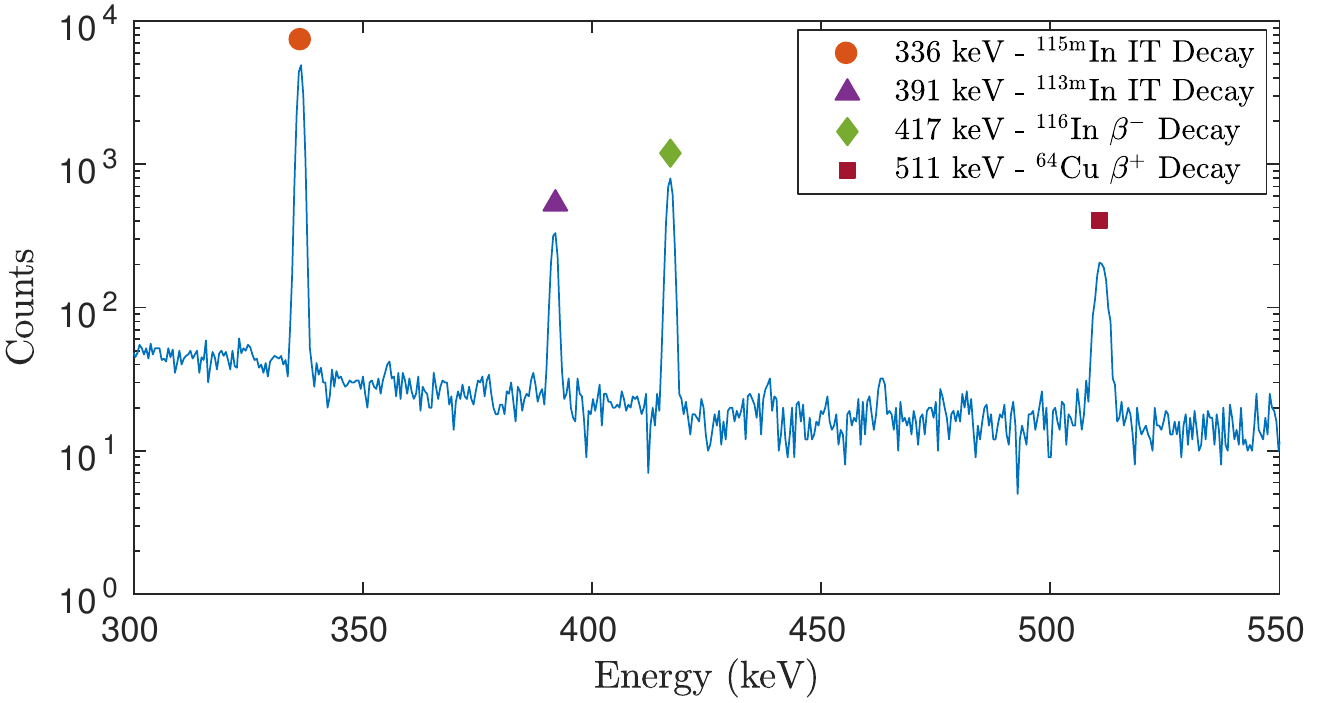}
          \refstepcounter{subfigure}       \label{fig:spectra_b}
    \end{subfigure}
    \caption{Example gamma spectra collected to monitor radioisotope production. (a)  Gamma spectrum for the \ce{^{47}Ti}(n,p)\ce{^{47}Sc} production pathway foils, counted using a LEPS detector and (b) gamma spectrum for the \ce{^{64}Zn}(n,p)\ce{^{64}Cu} production pathway foils, counted using an 80\% HPGe detector.}
     \label{fig:main_spectra}
\end{figure*}

For the  \ce{^{47}Sc} production experiments, the foils were counted simultaneously using a planar LEPS detector.
For the \ce{^{64}Cu} production experiments, the Indium foil was first counted separately using an 80\% HPGe detector, to capture the short-lived Indium activities.
This is due to the fact that the contaminant \ce{^{115}In}(n,$\gamma$) reaction results in the production of \ce{^{116m}In} which has a 54 minute half-life and results in the production of 1097 keV (58.5\% branching), 1293 keV (84.8\% branching) and 2112 keV (15.09\% branching) gamma-rays that in turn produce a significant number of 511 keV gammas from pair-production followed by annihilation \cite{Blachot2010}.
The foils were counted together again after approximately 4 hours of separate collection, to allow for nearly all of the produced \ce{^{116}In} to decay.
Example spectra for each production pathway can be seen in \autoref{fig:spectra_a} and \autoref{fig:spectra_b}.

\begin{figure*}
    \centering
    \begin{subfigure}[t]{0.49\textwidth}
        \centering
        \subfigimg[scale=0.6]{a)}{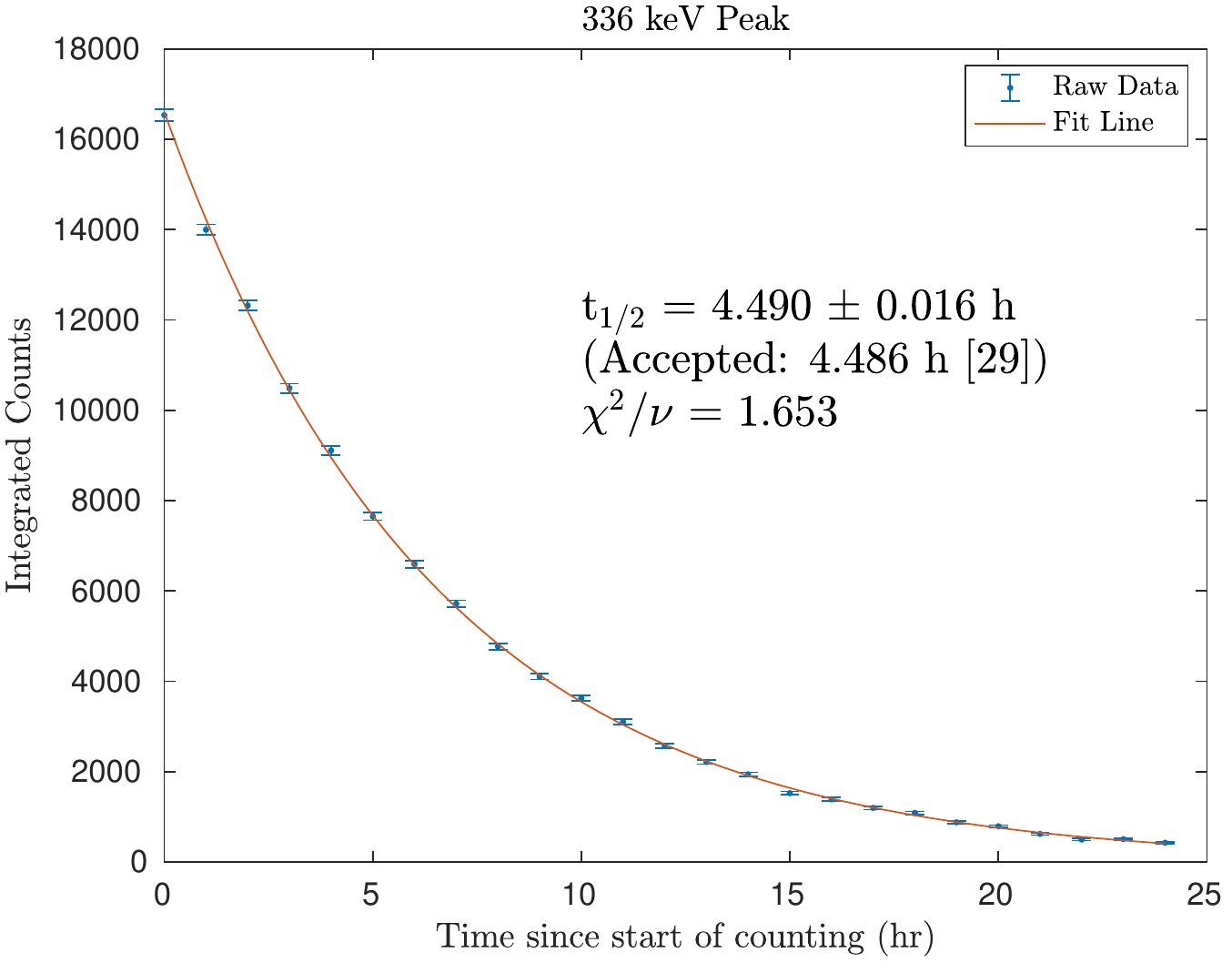}
         \refstepcounter{subfigure}\label{fig:decay_curve_336}
    \end{subfigure}%
     \begin{subfigure}[t]{0.49\textwidth}
        \centering
        \subfigimg[scale=0.6]{b)}{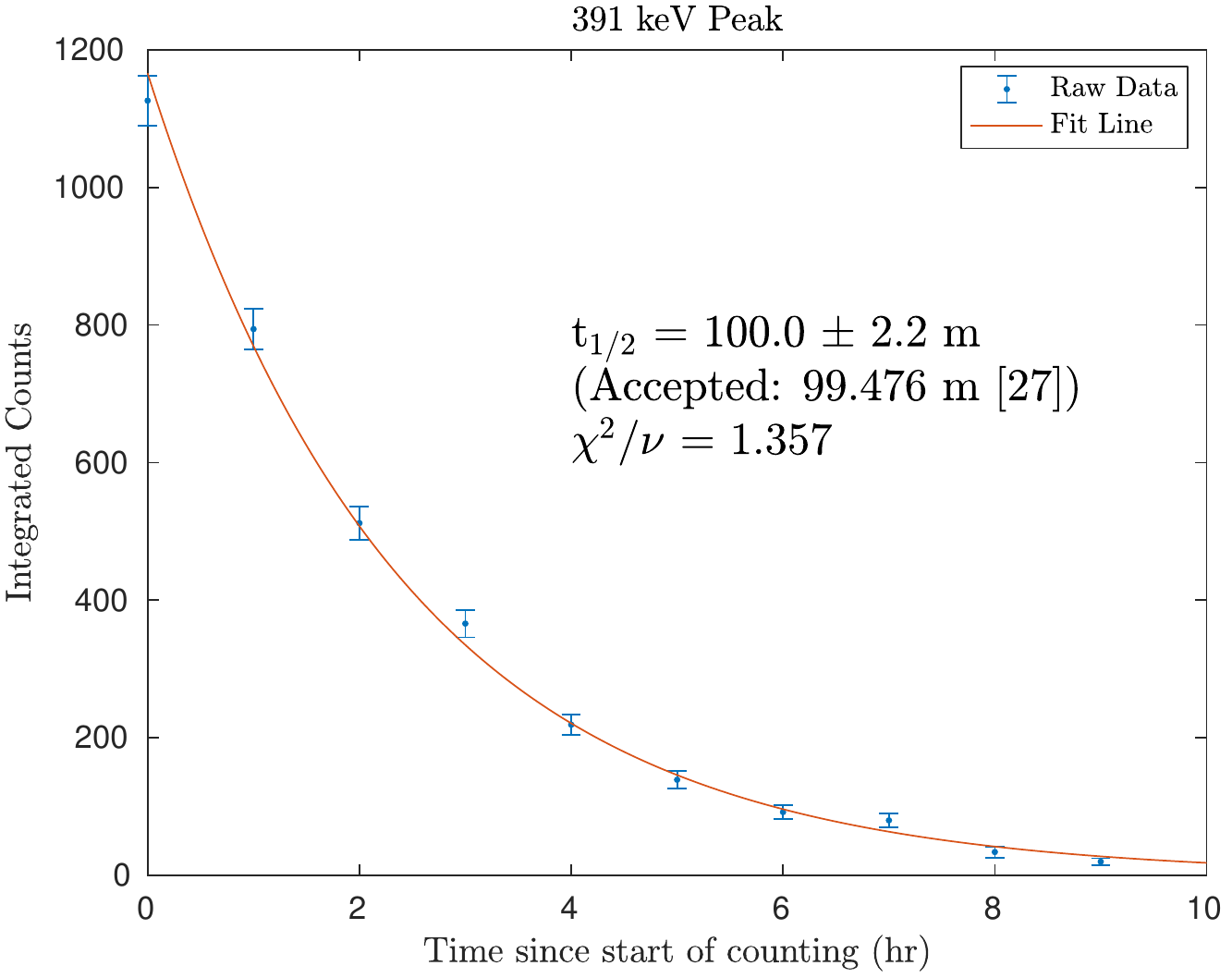}
         \refstepcounter{subfigure}\label{fig:decay_curve_391}
    \end{subfigure}%
    \\
    \begin{subfigure}[t]{0.49\textwidth}
        \centering
        \subfigimg[scale=0.6]{c)}{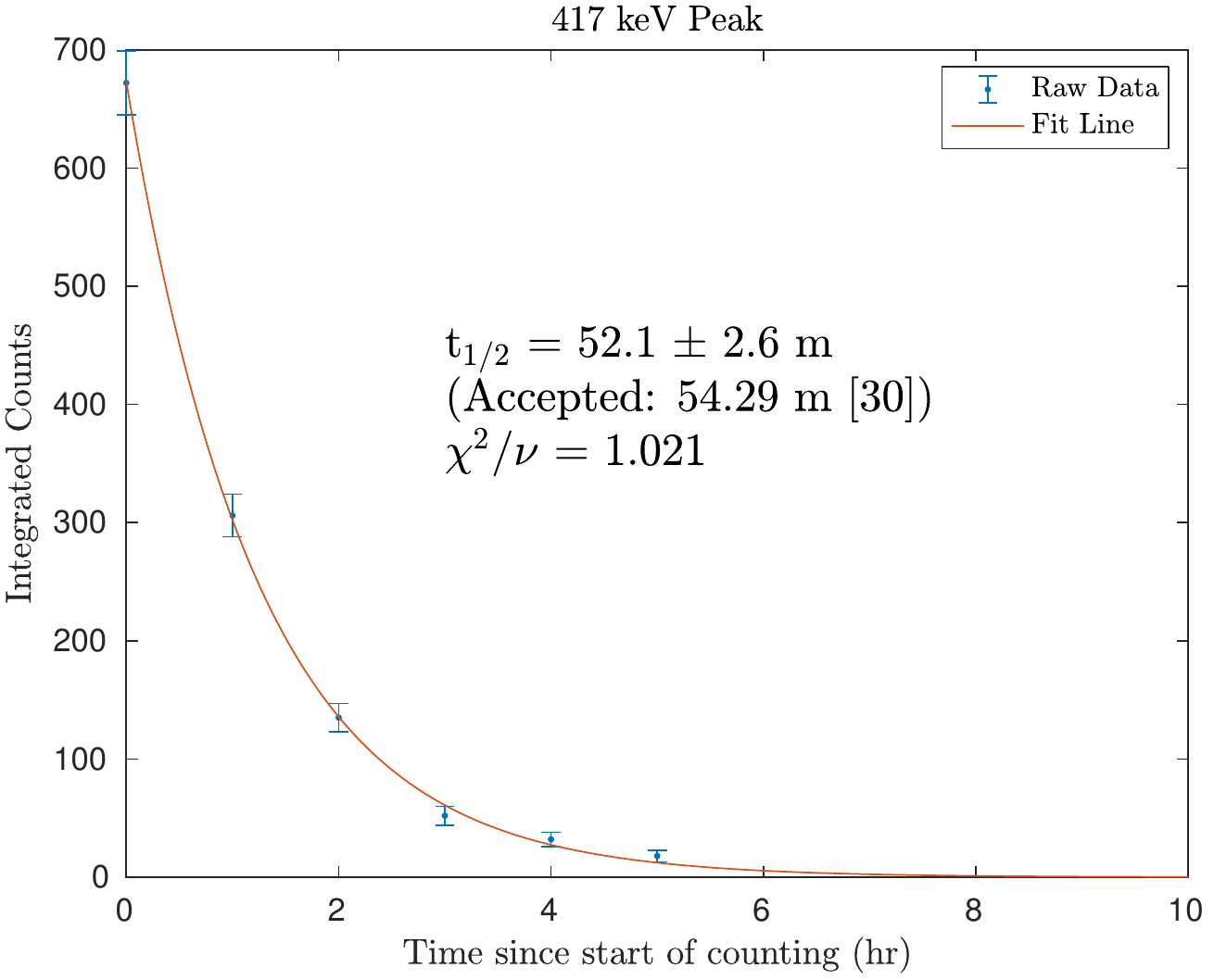}
                 \refstepcounter{subfigure}\label{fig:decay_curve_417}
    \end{subfigure}
     \begin{subfigure}[t]{0.49\textwidth}
        \centering
        \subfigimg[scale=0.6]{d)}{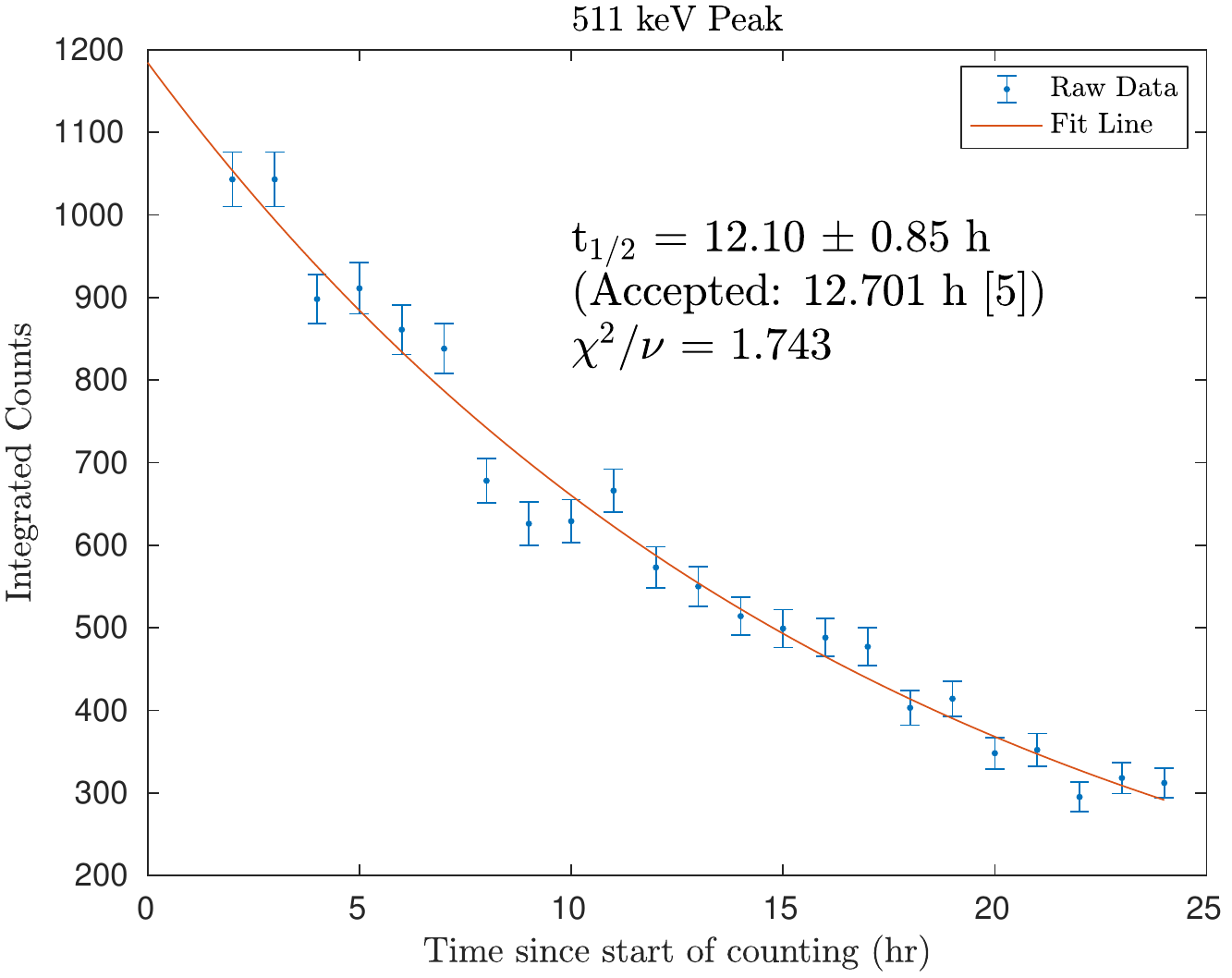}
        \refstepcounter{subfigure} \label{fig:decay_curve_511}
    \end{subfigure}%
    \caption{Decay curves used to verify photopeak transition assignment. (a) Decay curve for the isomeric transition of \ce{^{115m}In}, (b) decay curve for the isomeric transition of \ce{^{113m}In}, (c) decay curve for the $\beta^-$ decay of \ce{^{116}In}, and (d) decay curve for the $\beta^+$ decay of \ce{^{64}Cu}.}
     \label{fig:decay_curves}
\end{figure*}

To verify that each peak corresponds to the assigned decay product, spectra were acquired in a sequence of 15 - 30 minute intervals.
The resulting time series displayed in Figures \ref{fig:decay_curve_336} - \ref{fig:decay_curve_511} allow the fitting of exponential decay functions for each nuclide and comparison of the measured half-life with literature values.
The fitted functions for each transition agree (at the 1$\sigma$ confidence level) with accepted half-lives \cite{Burrows2007,Singh2007,Blachot2010a,Blachot2012,Blachot2010}, confirming the respective peak assignments.


The spectra for each sample were summed and the net peak areas were fitted using gf3, part of the RadWare analysis package from Oak Ridge National Laboratory   \cite{radford2000radware, Radford1995}.
The background-subtracted integrated counts in each photopeak, as well as the counting duration for each experiment, are tabulated in \autoref{tab:peak_counts}.

\begin{table*}
\centering
\caption{Counting times and photopeak counts for each of the (Zn/In) and (Ti/In)  experiments.  The uncertainties in photopeak counts are a combination of the fit error and counting statistics.}
\label{tab:peak_counts}
\resizebox{\textwidth}{!}{%
\begin{tabular}{lccccc}
\toprule
Reference Foil                     & \ce{^{nat}In}         & \ce{^{nat}In}       & \ce{^{nat}In} & \ce{^{nat}In}     & \ce{^{nat}In}  \\ \midrule
Reference Foil Mass (g)            & 0.248         & 0.248       & 0.241 & 0.247     & 0.248 \\ 
Target Foil                        & \ce{^{nat}Zn}         & \ce{^{nat}Zn}       & \ce{^{nat}Zn} & \ce{^{nat}Ti}     & \ce{^{nat}Ti}  \\ 
Target Foil Mass (g)               & 0.538         & 0.521       & 0.542 & 0.337     & 0.337   \\ 
Irradiation Time, $t_i$ (s)           & 10800         & 10800       & 12629 & 11837     & 14254   \\ 
Delay Time, $t_d$ (s)             & 1785         & 16185       & 2290 & 89408    & 2390   \\ 
Counting Time, $t_c$ (s)               & 91188        & 54008       & 54002 & 86424    & 93631   \\ 
Photopeak Counts, 336 keV (\ce{^{115m}In}) & 113665 $\pm$ 1490 & 76321 $\pm$ 275 & 39895 $\pm$ 201 & 2122 $\pm$ 55 & 55102 $\pm$ 268   \\ 
Photopeak Counts, 391 keV (\ce{^{113m}In}) & 3382 $\pm$ 171    & 890 $\pm$ 40    & 3505 $\pm$ 54 &    \hrulefill       &  \hrulefill      \\ 
Photopeak Counts, 511 keV (\ce{^{64}Cu})   & 16055 $\pm$ 643   & 12852 $\pm$ 118 & 27164 $\pm$ 159 &    \hrulefill       &   \hrulefill     \\ 
Photopeak Counts, 159 keV (\ce{^{47}Sc})   &   \hrulefill            &    \hrulefill         &   \hrulefill    & 3877 $\pm$ 83 & 5544 $\pm$ 257 \\ \bottomrule
\end{tabular}%
}
\end{table*}

\subsection{Experimental verification of incident neutron energy}\label{sec:lee_sec}

\begin{figure}
 \centering
 \includegraphics[scale=0.55]{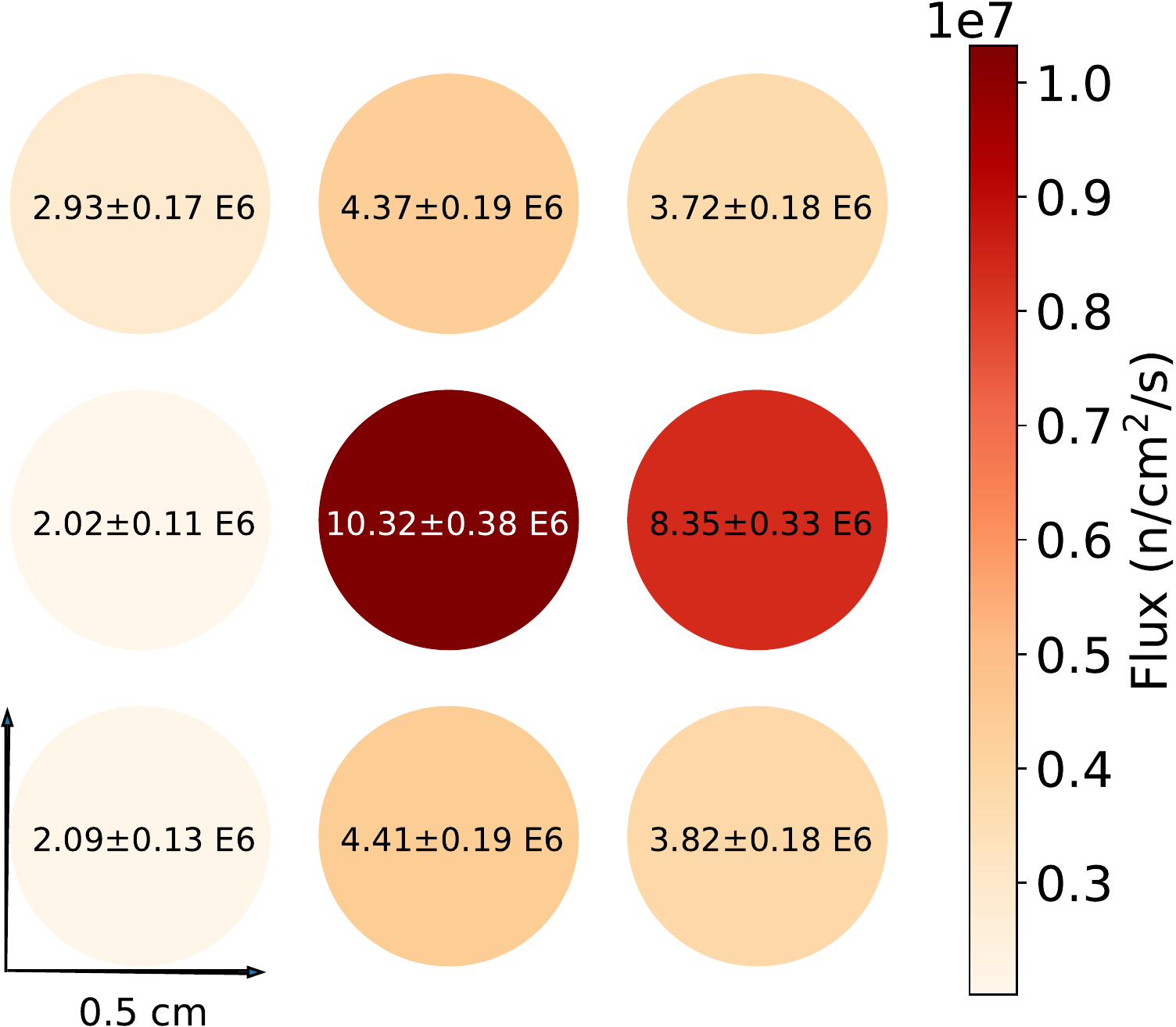}
 \caption{Relative fluxes as seen by a 3 x 3 array of indium foils. The central foil corresponds to the location in which target and monitor foils were mounted during the cross section measurements, verifying that the beam is centered on the middle of mounted foils.}
 \label{fig:in_heat_plot}
\end{figure}

As shown in \autoref{sec:sample_loading} above, the effective neutron energy depends on the angle range subtended by the sample with respect to the incident deuteron beam.
 In order to determine this angle it is necessary to measure the lateral location of the beam with respect to the sample location.
 This centroid position of the beam was measured using a 3 x 3 array of 0.5 cm diameter indium foils.
 The relative activity of these foils was then determined via post-irradiation counting of the \ce{^{115m}In} isomer ($t_{1/2}$ = 4.486 h) \cite{Blachot2012}.
 \autoref{fig:in_heat_plot}  shows the measured activities for these 9 indium foils.
 Based on these values we are able to verify that the beam was indeed vertically centered on the middle of the zinc and titanium samples,  with a slight asymmetry of the neutron flux in the horizontal direction, accounted for in MCNP6 modeling of the energy-differential neutron flux. This small asymmetry likely contributes to the effective energy bin being lower than the 2.78 MeV expected for 0\degree\ neutron emission angle in \autoref{fig:scatt_angle}.

\subsection{Calculation of measured cross sections}\label{sec:calcs_sec}

For a thin target consisting of \(N_T\) target nuclei (with a reaction cross section $\sigma\pp{\bar{E}}$), subjected to a constant neutron flux $\phi\pp{\bar{E}}$, the rate of production ($R$) of the product nucleus will be:

\begin{equation}
R = N_T \sigma\pp{\bar{E}} \phi\pp{\bar{E}} 
\end{equation}

If the target is subjected to this flux  for an irradiation  time $t_i$ and decays for a  delay time $t_d$ (after end-of-beam) before gamma ray spectrum acquisition occurs for a counting time  $t_c$, then the number of product decays ($N_D$; with decay constant $\lambda$) during the acquisition will be:

\begin{align}
N_D &= \dfrac{R}{\lambda}\pp{1 - e^{-\lambda t_i}} e^{-\lambda t_d} \pp{1 - e^{-\lambda t_c}}
\\
&= \dfrac{N_T \sigma\pp{\bar{E}} \phi\pp{\bar{E}} }{\lambda}\pp{1 - e^{-\lambda t_i}} e^{-\lambda t_d} \pp{1 - e^{-\lambda t_c}} \nonumber
\end{align}


If this decay emits a gamma ray with absolute intensity $I_\gamma$ (photons emitted per decay), and is detected with an absolute efficiency  of $\epsilon_\gamma$ (photons detected / photons emitted), then the number of observed gamma rays during the acquisition will be:


\begin{align}
N_{\gamma} &= N_D \epsilon_\gamma I_\gamma \\
&=  \epsilon_\gamma I_\gamma  \dfrac{N_T \sigma\pp{\bar{E}} \phi\pp{\bar{E}} }{\lambda}\pp{1 - e^{-\lambda t_i}}  e^{-\lambda t_d} \pp{1 - e^{-\lambda t_c}} \nonumber
\end{align}

Solving this equation for the  cross section results in:

\begin{align}\label{eqn:single_xs_eqn}
\sigma\pp{\bar{E}} = \dfrac{N_{\gamma}\lambda}{N_T \epsilon_\gamma I_\gamma  \phi\pp{\bar{E}}  \pp{1 - e^{-\lambda t_i}} e^{-\lambda t_d} \pp{1 - e^{-\lambda t_c}}}
\end{align}


\autoref{eqn:single_xs_eqn} can be used to determine the unknown (n,p) cross sections relative to the well-known \ce{^{115}In}(n,n')\ce{^{115m}In} and \ce{^{113}In}(n,n')\ce{^{113m}In} inelastic scattering cross sections since the Zn and Ti samples were co-irradiated with indium foils.
This approach has a number of advantages since the result is independent of neutron flux and only depends on the relative detector efficiencies at each gamma-ray energy.
 \autoref{eqn:calc_eqn} shows the ratio of the cross sections determined using this approach, in which subscript $P$ indicates a quantity for either \ce{^{64}Cu} or \ce{^{47}Sc}, and subscript $In$ indicates a quantity for either the \ce{^{113m}In} or \ce{^{115m}In} isomer.
A minor term was added to correct for the small self-attenuation of the gamma rays emitted by the activated foils:

\begin{align}\label{eqn:calc_eqn}
\dfrac{\sigma_P}{\sigma_{In}} &=  \dfrac{N_{\gamma,P}}{N_{\gamma,In}}  \dfrac{N_{T,In}}{N_{T,P}} \dfrac{\lambda_P}{\lambda_{In}} \pp{\dfrac{1 - e^{-\lambda_{In}t_i}}{1 - e^{-\lambda_{P}t_i}}} \dfrac{e^{-\lambda_{In}t_d}}{ e^{-\lambda_{P}t_d}} \times \\
&\times \pp{\dfrac{1 - e^{-\lambda_{In}t_c}}{1 - e^{-\lambda_{P}t_c}}} \dfrac{\epsilon_{In}}{\epsilon_P}  \dfrac{I_{\gamma,In}}{I_{\gamma,P}} \dfrac{e^{-\mu_{In}x_{In}/2}\times e^{-\mu_{In}x_{P}}}{e^{-\mu_{P}x_{P}/2}} \nonumber
\end{align}

where:

\begin{itemize}

\item $N_{\gamma}$ is the integrated counts under a photopeak,

\item $\sigma$ is the cross section for either the production of a product or isomer [mb],

\item $N_T$ is the initial number of target nuclei,

\item $\lambda$  is the decay constant [s$^{-1}$],

\item $t_i$ is the irradiation time [s],

\item $t_d$ is the delay time (between the end-of-beam and the start of counting) [s],

\item $t_c$ is the counting time   [s],

\item $\epsilon$ is the  detector efficiency for a particular photopeak,

\item $I_\gamma$ is the decay gamma ray absolute intensity [\%],

\item $\mu$ is the photon attenuation coefficient for a particular decay gamma ray in a foil [cm$^{-1}$],

\item and $x$ is the thickness of foil traversed by a particular decay gamma ray [cm] 
\end{itemize}

In addition to the \ce{^{115}In}(n,n')\ce{^{115m}In} reference cross section, the \ce{^{115}In}(n,$\gamma$)\ce{^{116m}In}  ($t_{1/2}$ = 54.29 min \cite{Blachot2010}) activity can be used to determine the  \ce{^{64}Zn}(n,p) and \ce{^{47}Ti}(n,p) cross section.
 The capture activity is potentially subject to contamination from lower energy, especially thermal, \enquote{room return} neutrons since the (n,$\gamma$) cross section at 25 meV is approximately 2,000 times greater than at 2.7 MeV  \cite{Capote2012,zsolnay2012technical}.


With the exception of decay constants  and time measurement, which have negligible uncertainty compared to other sources of uncertainties in this work, each of the parameters in this model carries an uncertainty.
Based on the assumption that these uncertainties are uncorrelated, the total relative statistical uncertainty $\delta_\sigma$ is calculated by taking the quadrature sum of the relative uncertainties of each parameter  $\delta_i$:

\begin{equation}
\delta_\sigma = \norm{ \ensuremath{\vec{\delta}} }_2 = \sqrt{\sum_{i=1}^N  \delta_i^2  }
\end{equation}

This total  uncertainty is plotted as the cross section uncertainty in \autoref{fig:sc_xs} and \autoref{fig:cu_xs_zoom}.

\subsection{Systematic uncertainties}

The largest source of systematic uncertainty in the cross section determined via the \enquote{ratio approach} is the 2.586\% uncertainty in the \ce{^{115}In}(n,n')\ce{^{115m}In}  cross section and the 1.447\% uncertainty in the \ce{^{113}In}(n,n')\ce{^{113m}In}  cross section  \cite{Capote2012,zsolnay2012technical}.
 An additional uncertainty arises from the fact that the Zn/Ti samples are not located at exactly the same location as the indium monitor foils, and are therefore not subject to precisely the same neutron flux.
 However, the MCNP6 simulations shown in \autoref{fig:mcnp_flux} indicate that the difference in the flux that the two foils are subjected to is less than 1\%, negligible compared to other sources of systematic uncertainty.
 Other monitor foils could be used instead of indium, with \ce{^{58}Ni}(n,p)\ce{^{58}Co} (\ce{^{58}Co} $t_{1/2}$ = 70.86 d \cite{Nesaraja2010}) being one possible candidate, but the 4.486 hour and 99.476 minute half-lives of the \ce{^{115m}In} and \ce{^{113m}In} isomers \cite{Blachot2012,Blachot2010a}, respectively, make indium a better candidate for measuring the production of radionuclides with lifetimes much less than 71 days.
The largest source of uncertainty in energy window arises from uncertainties in the actual dimension of the deuteron beam on the production target.
 We believe, based on \enquote{burn marks} on the neutron production target, that the beam was approximately circular, with a flat intensity profile and a 5 mm diameter.
 However, every 1 mm change in the beam radius would cause a 0.028 MeV shift in the centroid and a 0.053 MeV increase in the effective energy bin width, which places a natural limit on the reported effective neutron energy.

A much smaller systematic uncertainty arises from the fact that the two (n,p) cross sections and the reference In(n,n') cross sections have slightly different thresholds.  
The total activity in the In produced by the low energy neutrons (below the \enquote{knee} near 2.25 MeV in  \autoref{fig:mcnp_flux}) is 2.17\%.  
The corresponding values from TALYS for the $^{64}$Cu and $^{47}$Sc activity are 0.24\% and 0.85\%, respectively.  
If we assume an uncertainty of $\pm$25\% in the TALYS calculations in this energy region it would introduce an additional systematic uncertainty in the $^{+10}_{-20}$ keV effective energy bin of $\pm$1.6 keV for $^{64}$Cu and $\pm$5.7 keV for $^{47}$Sc. 
As these are smaller than the precision of the existing effective energy bin, they can be considered negligible.

\section{Results}

Using the ratio method described, the cross sections for the \ce{^{47}Ti}(n,p)\ce{^{47}Sc} and \ce{^{64}Zn}(n,p)\ce{^{64}Cu} reactions have been calculated for an incident neutron energy of $E_n$ =2.76$^{+0.01}_{-0.02}$ MeV.
These values are recorded in \autoref{tab:xs_results}.


%

\begin{table}
\centering
\caption{Results of cross section measurement. Note that the last data point for the \ce{^{47}Sc} measurement (marked with *) was performed at a slightly different beam spot location, leading to a difference in effective neutron energy.}
\label{tab:xs_results}
\begin{tabular}{@{}cS[table-format=+2.1,
                  table-figures-uncertainty=1]@{}}
\toprule
Reaction                   & \multirows{1}*{\minitab[c]{$\sigma$($E_n$ = 2.76$^{+0.01}_{-0.02}$ MeV) (mb)}}       \\ \midrule
\multirows{3}*{\minitab[c]{\ce{^{64}Zn}(n,p)\ce{^{64}Cu} \\ (relative to \ce{^{113}In})}} & 49.9 \pm 3.2 \\ 
											& 49.2 \pm 2.7 \\ 
											& 49.0 \pm 2.5 \\  \\[0.1ex]  
\multirows{3}*{\minitab[c]{\ce{^{64}Zn}(n,p)\ce{^{64}Cu} \\ (relative to \ce{^{115}In})}} & 45.9 \pm 2.6 \\ 
											& 46.5 \pm 1.7 \\ 
											& 46.8 \pm 3.2  \\ \\[0.1ex]    
\multirows{2}*{\minitab[c]{\ce{^{47}Ti}(n,p)\ce{^{47}Sc} \\ (relative to \ce{^{115}In})}} & 25.9 \pm 1.2 \\ 
											& 26.7 \pm 1.4*                   \\ \bottomrule
\end{tabular}
\end{table}

\begin{figure}
 \centering
 \includegraphics[scale=0.65]{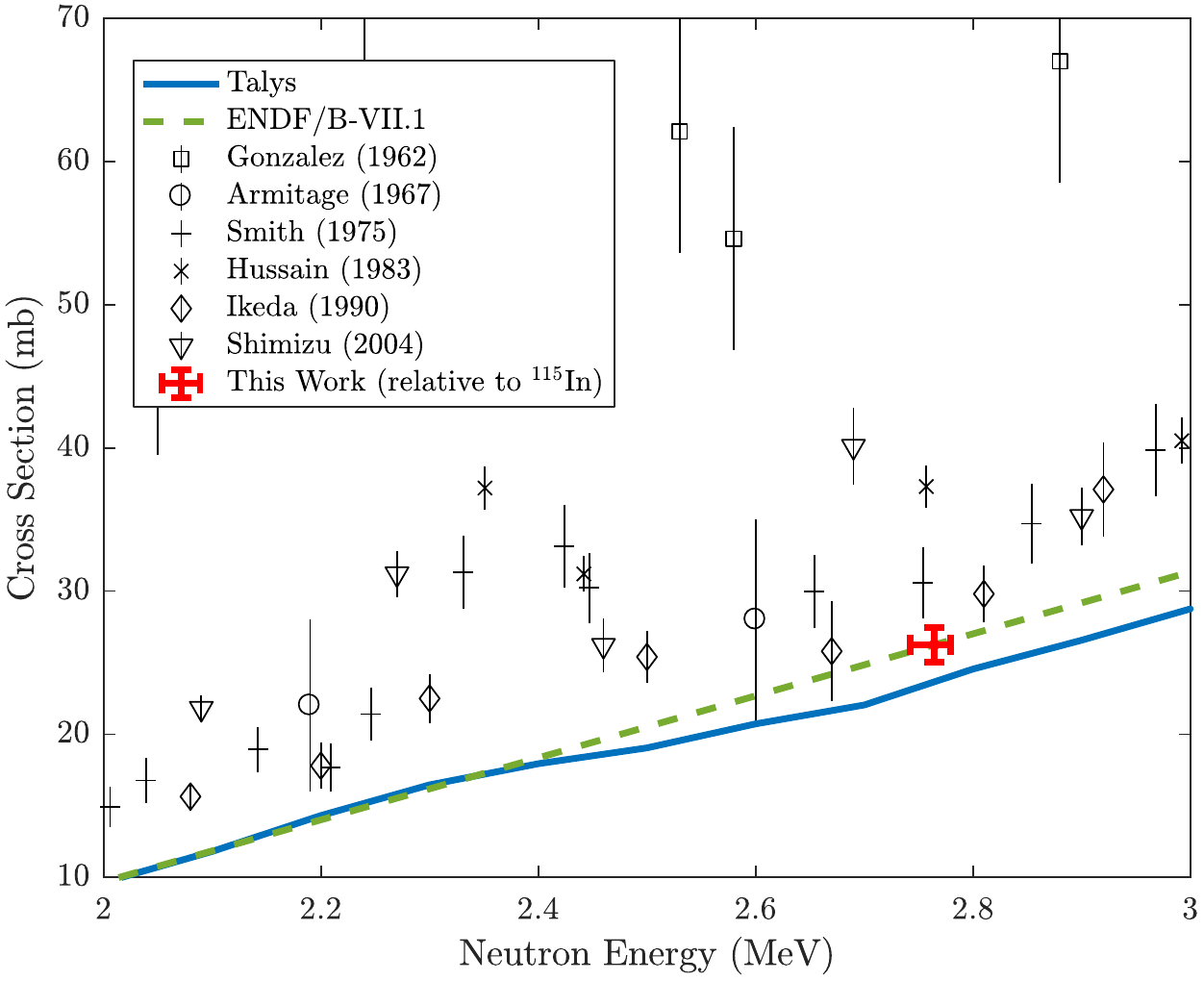}
 \caption{Measured \ce{^{47}Ti}(n,p)\ce{^{47}Sc} cross section relative to indium activation.}
 \label{fig:sc_xs}
\end{figure}
\begin{figure}
 \centering
 \includegraphics[scale=0.65]{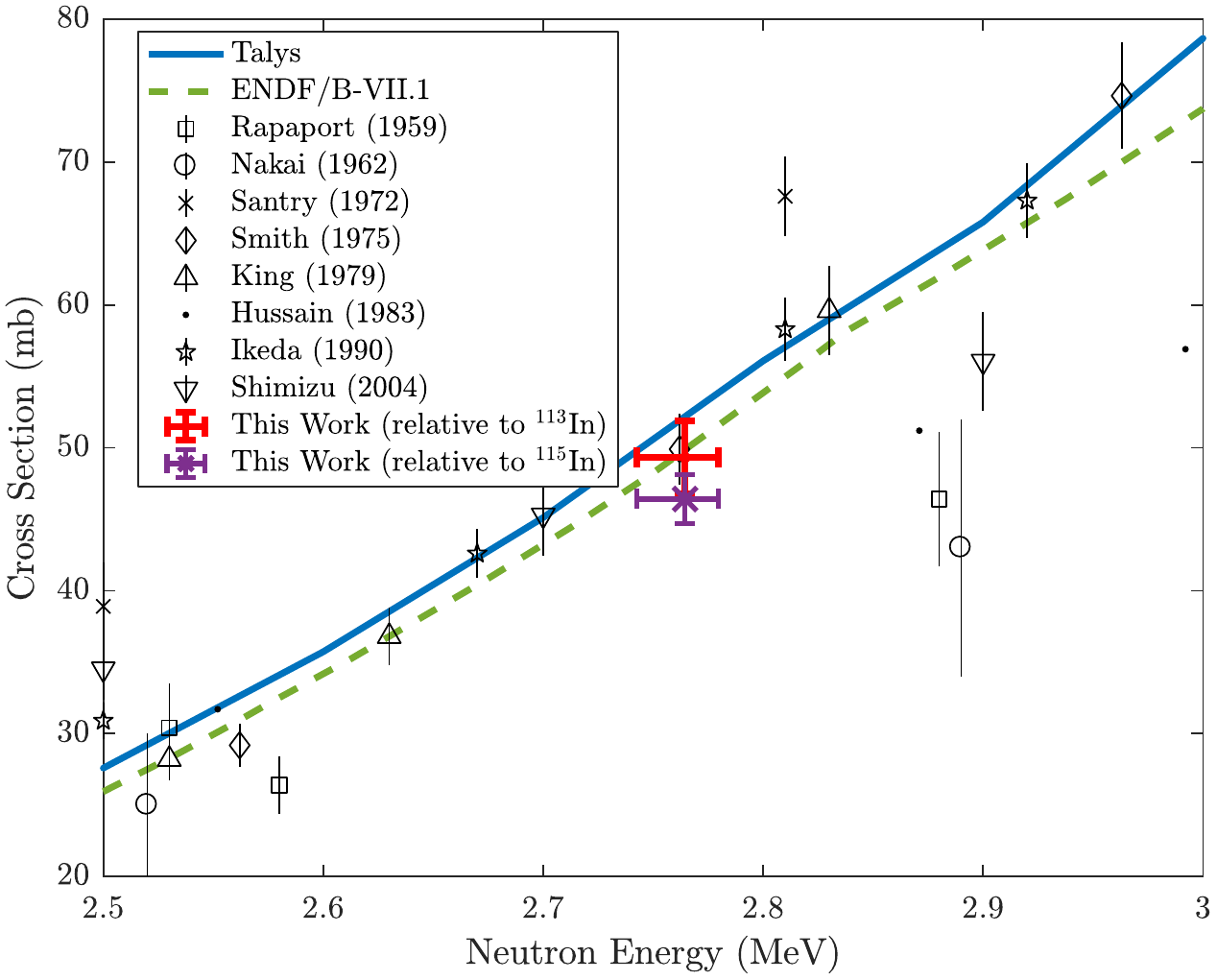}
 \caption{Measured \ce{^{64}Zn}(n,p)\ce{^{64}Cu} cross section relative to indium activation.}
 \label{fig:cu_xs_zoom}
\end{figure}

Figures \ref{fig:sc_xs} and \ref{fig:cu_xs_zoom} present the determined cross sections for the production of  \ce{^{47}Ti}(n,p)\ce{^{47}Sc} and \ce{^{64}Zn}(n,p)\ce{^{64}Cu}  relative to literature data retrieved from EXFOR \cite{Otuka2014272,PhysRev.114.565,doi:10.1143/JPSJ.17.1215,paulsen1967cross,doi:10.1139/p72-336,Smith1975,King1979,Hussain1983,Ikeda1991,Shimizu2004543,PhysRev.126.271,Armitage1967,Ikeda1990,Senga2000,Shimizu2004975}.
The weighted average of the measurements give   49.3 $\pm$ 2.6 mb (relative to \ce{^{113}In}) and 46.4 $\pm$ 1.7 mb (relative to \ce{^{115}In})   for \ce{^{64}Zn}(n,p)\ce{^{64}Cu},   and 26.26 $\pm$  0.82 mb for  \ce{^{47}Ti}(n,p)\ce{^{47}Sc}.
The \ce{^{64}Zn}(n,p)\ce{^{64}Cu} cross section measured in this work is consistent with  other literature results, but with a smaller uncertainty (\textless 5\%).
However, in the case of the \ce{^{47}Ti}(n,p)\ce{^{47}Sc} cross section, our results are consistent with the results from the Smith (1975), Armitage (1967), and Ikeda (1990) groups \cite{Smith1975,Armitage1967,Ikeda1990} and both the ENDF/B-VII.1 \cite{Chadwick2011} and TALYS \cite{Koning2012}  values, but significantly below the results from the Hussain (1983), Gonzalez (1962), and Shimizu (2004) groups \cite{Hussain1983,PhysRev.126.271,Shimizu2004975}.






As mentioned above, the cross section can be obtained relative to both the inelastic scattering cross sections on \ce{^{113}In} and \ce{^{115}In}, and the capture of fast, unmoderated neutrons on \ce{^{115}In}.
The result for the production of \ce{^{116}In} via the \ce{^{115}In}(n,$\gamma$) reaction was shown to be consistent with activation predominantly from the capture of fast neutrons, rather than from \enquote{room return} thermal neutrons.
The MCNP neutron spectrum in   \autoref{fig:mcnp_flux} confirms this - thermal and epithermal neutrons make up only 0.0771\% of the total neutron population.
 This will be discussed in greater detail in the conclusion section below.


\section{Discussion}

The proximity of the target to the neutron production surface opens the possibility of performing a measurement of the cross section over a limited energy range via mounting the samples slightly off-axis with respect to the beam.
 This could be accomplished using the 9-foil sample holder described in \autoref{sec:lee_sec} above.
 Mounting samples at each of these positions would subject the samples to neutrons with energies ranging from 2.765 MeV at the central location to 2.616 MeV at the four corners, with the other locations having intermediate energy values.
 These sorts of multi-sample measurements could be used to determine the \enquote{rising edge} of the cross sections, aiding in the development of optical models for the reactants.

These measurements also highlight the possibility of using fast neutrons from DD and/or DT generators to produce meaningful quantities of radioisotopes for a wide range of applications via charge exchange reactions, such as (n,p) and (n,$\alpha$).
 Many applications, including diagnostic and therapeutic medical use, require mCi activity levels.
 For the production of a radionuclide sample, the saturation activity ($A_{saturation}$) is achieved at secular equilibrium:

 \begin{equation}
R_{production} = R_{decay} = \lambda N_{product}
\end{equation}

While the saturation activity represents the maximum activity that can be made at a generator with a given total neutron output, there may be situations where either a smaller activity is needed, or a shorter irradiation is desired. 
In this case, it is useful to introduce a neutron utilization factor ($\eta_x$).  $\eta_x$ is the constant of proportionality between $R_n$, the neutron source output (in neutrons/second), and the saturation activity:

\begin{equation}
A_{saturation} = \eta_x R_n
\end{equation}

$\eta_x$ represents the likelihood that a neutron produced in the generator will create $x$, the isotope of interest. 
It includes the overlap between the production target and the locus where the neutrons are being created, and the fraction of nuclear reactions which generate the desired activity $x$:

\begin{align}
\eta_x &= \dfrac{1}{R_n} \int\displaylimits_{production\ target} \phi\pp{\mathbf{r}}\bar{\sigma}_x\ \rho_{target}\pp{\mathbf{r}}\text{d}\mathbf{V}, \nonumber\\
\text{d}\mathbf{V} &= \mathbf{r}^2 \text{d}\mathbf{r} \sin{\theta}\, \text{d}\theta\, \text{d}\varphi 
\end{align}

where $\bar{\sigma}_x$ is the average cross section producing the radionuclide of interest, $\rho_{target}\pp{\mathbf{r}}$ is the density of the target as a function of position, and $\phi\pp{\mathbf{r}}$ is the neutron flux (in n/cm$^2$/s) as a function of position.   $\eta_x$ allows us to cast the activity produced in a given irradiation time $t_i$ as:

\begin{equation}
A\pp{t_i} = \eta_x R_n \pp{1-e^{-\lambda t_i}}
\end{equation}

Maximizing $\eta_x$ would be the goal of any engineering design to produce a desired activity using a neutron generator at a minimum of cost and radiological impact.

%
%
%
%
%
%
%

An optimal design for the neutron generator would also allow for the possibility of reflecting fast neutrons back onto the target to maximize their utilization for radionuclide production.
 This sort of \enquote{flux trap} has been used for the production of radionuclides in reactors, but has not to date been optimized for use with fast neutrons  at DD and/or DT neutron sources.
 The HFNG, with its self-loading target and \enquote{flux trap} geometry, has many features that make it well-suited for such isotope production purpose.
 Switching to DT operation would dramatically increase the flux as well as the production cross section, since (n,p) tends to be significantly larger at 14 MeV.
 However, the higher neutron energy would also open the (n,pn) channels.
 In the case of \ce{^{47}Sc}, this would lead to the presence of \ce{^{46}Sc} ($t_{1/2}$ = 83.79 d \cite{Wu2000}) in the sample, which might pose some concerns for medical applications.
 However, this is not an issue for \ce{^{64}Cu} since the (n,pn) channel leads to the production of stable \ce{^{63}Cu}.
 
Assuming a neutron flux of  $\sci{1.3}{7}$ neutrons/cm$^2$s on the target, masses of 0.533 g of natural zinc and 0.337 g of natural titanium, and cross sections of 47.5 mb for \ce{^{64}Zn}(n,p)\ce{^{64}Cu}   and 26.26  mb for  \ce{^{47}Ti}(n,p)\ce{^{47}Sc}, theoretical saturation activities for current operation at the time of this work are estimated to be 1.5 kBq of \ce{^{64}Cu} and 0.11 kBq of \ce{^{47}Sc}.
This falls short of the mCi (37 MBq) level required for commercial application by a factor of 3-4 orders of magnitude, but with the operation of the second deuterium ion source, increased current, and fast neutron reflection, this goal may  well be within reach. 
By increasing the activation target thickness to 1 cm (a factor of 10), switching to DT operation (a factor of 80), increasing current and running the second ion source (a factor of 60), and relying upon the higher (n,p) cross section at DT energies (a factor of approximately 3), we believe saturation activities of approximately 6 mCi of \ce{^{64}Cu} and 0.5 mCi of \ce{^{47}Sc} can be achieved.  
The activities produced at the end of irradiation averaged 453.8 Bq of \ce{^{64}Cu}, and 31.6 Bq of \ce{^{47}Sc}.
Assuming a conservative neutron source output of $10^8$ neutrons / second, we can estimate that, in present operation, the HFNG has an average $\eta_{64Cu} \approx \sci{3.0}{-5}$ for \ce{^{64}Cu} and $\eta_{47Sc} \approx \sci{1.1}{-5}$ for  \ce{^{47}Sc}. 
This falls approximately 4 orders of magnitude short of the $\eta_x \approx 0.37$ needed for mCi-scale production.
A factor of 10 in $\eta_x$ could easily be gained through use of targets  1-cm in thickness without worry of contaminating reaction channels opening up, but $\eta_x$ gains beyond this will require modification of operation conditions.

 \section{Conclusion and Future Work}

Using activation methods on thin foils, the \ce{^{47}Ti}(n,p)\ce{^{47}Sc} and \ce{^{64}Zn}(n,p)\ce{^{64}Cu} production cross sections were measured for  2.76$^{+0.01}_{-0.02}$ MeV neutrons produced using the High Flux Neutron Generator (HFNG) at UC Berkeley.
The cross sections were measured with less than  5\% uncertainty relative to the well-known \ce{^{115}In}(n,n')\ce{^{115m}In} and \ce{^{113}In}(n,n')\ce{^{113m}In} fast neutron cross sections \cite{Capote2012,zsolnay2012technical}.
The measured values of  26.26 $\pm$  0.82 mb and  49.3 $\pm$ 2.6 mb (relative to \ce{^{113}In}) or 46.4 $\pm$ 1.7 mb (relative to \ce{^{115}In}), respectively, are consistent with earlier experimental data and theoretical models, but have smaller uncertainties than previous measurements.
 
 In addition, the   production of the \ce{^{116}In}  via the \ce{^{115}In}(n,$\gamma$) reaction was close to the value one would expect given an effective incident neutron energy of 2.45 MeV.  
 While this is not consistent with the average neutron energy at the target location (2.76$^{+0.01}_{-0.02}$ MeV), the fact that it was close indicates the paucity of thermal neutrons in this central location.  
 This in turn highlights the usefulness of such compact DD-neutron sources  for producing \enquote{clean} activities via the (n,p) channel.  
The use of DD neutron generators can be an efficient method for the measurement of low-energy (n,p) reaction channels, as well as a relative method used to normalize measurements at higher neutron energies.
 In addition to improving the value of these measurements for nuclear reaction evaluation, our results highlight the potential use of compact neutron generators for the production of radionuclides locally for medical applications.

 It is worth noting that at the time of publication, the  HFNG is now operating at close to 10$^9$ n/sec, with
a clear path towards 10$^{10}$.
 Future work will involve the continued measurement of the (n,p) production cross sections for various other emerging therapeutic and diagnostic radioisotopes, to expand the toolset of options available for modern medical imaging and cancer therapy.
This will focus on radionuclides which permit more customized and precise dose deposition, as well as patient-specific treatments.
 

 \section{Acknowledgements}
 
 We would like to particularly point out the crucial role played by Cory Waltz in the design and commissioning of the HFNG.
 We acknowledge Glenn Jones of G\&J Jones Enterprises of Dublin, CA for the construction  of the High Flux Neutron Generator. 
 Lastly, we would like to acknowledge the students in the Nuclear Reactions and Radiation (NE102) laboratory course at UC Berkeley who participated in these experiments, including Joe Corvino, Nizelle Fajardo, Scott Parker and Evan Still.  
 
 This work has been carried out at the University of California, Berkeley, and performed under the auspices of the U.S. Department of Energy by Lawrence Livermore National Laboratory under contract \# DE-AC52-07NA27344 and Lawrence Berkeley National Laboratory under contract \# DE-AC02-05CH11231.
Funding has been provided from the US Nuclear Regulatory Commission, the US Nuclear Data Program, the Berkeley Geochronology Center, NSF ARRA Grant \# EAR-0960138, the University of California Laboratory Fees Research Grant \# 12-LR-238745, and  DFG Research Fellowship \# RU 2065/1-1.

\bibliographystyle{elsarticle-num}
\bibliography{./library}





















\end{document}